\documentclass[preprint2]{aastex}

\usepackage{amsmath}
\usepackage{graphicx}
\usepackage{epsfig,psfrag,epic,eepic}
\usepackage{textcomp}
\usepackage{times}
\usepackage{longtable}
\usepackage{lscape}

\slugcomment{Accepted to the Astrophysical Journal}

\shorttitle{Late M-dwarf Multiplicity}
\shortauthors{Janson et al.}

\begin{document}

\title{The AstraLux Multiplicity Survey: Extension to Late M-dwarfs\altaffilmark{*}}

\author{Markus Janson\altaffilmark{1,2}, 
Carolina Bergfors\altaffilmark{2,3}, 
Wolfgang Brandner\altaffilmark{2},
Natalia Kudryavtseva\altaffilmark{2,4},
Felix Hormuth\altaffilmark{2}, 
Stefan Hippler\altaffilmark{2}, 
Thomas Henning\altaffilmark{2}
}

\altaffiltext{*}{Based on observations collected at the Centro Astron\'omico Hispano Alem\'an (CAHA) at Calar Alto, operated jointly by the Max-Planck Institute for Astronomy and the Instituto de Astrof\'isica de Andaluc\'ia (CSIC).}
\altaffiltext{1}{Queen's University Belfast, Belfast, Northern Ireland, UK; \texttt{m.janson@qub.ac.uk}}
\altaffiltext{2}{Max Planck Institute for Astronomy, Heidelberg, Germany}
\altaffiltext{3}{University College London, London, UK}
\altaffiltext{4}{Astronomisches Rechnen-Institut, Heidelberg, Germany}

\begin{abstract}\noindent
The distribution of multiplicity among low-mass stars is a key issue to understanding the formation of stars and brown dwarfs, and recent surveys have yielded large enough samples of nearby low-mass stars to study this issue statistically to good accuracy. Previously, we have presented a multiplicity study of $\sim$700 early/mid M-type stars observed with the AstraLux high-resolution Lucky Imaging cameras. Here, we extend the study of multiplicity in M-type stars through studying 286 nearby mid/late M-type stars, bridging the gap between our previous study and multiplicity studies of brown dwarfs. Most of the targets have been observed more than once, allowing us to assess common proper motion to confirm companionship. We detect 68 confirmed or probable companions in 66 systems, of which 41 were previously undiscovered. Detections are made down to the resolution limit of $\sim$100~mas of the instrument. The raw multiplicity in the AstraLux sensitivity range is 17.9\%, leading to a total multiplicity fraction of 21--27\% depending on the mass ratio distribution, which is consistent with being flat down to mass ratios of $\sim$0.4, but cannot be stringently constrained below this value. The semi-major axis distribution is well represented by a log-normal function with $\mu_{\rm a} = 0.78$ and $\sigma_{\rm a} = 0.47$, which is narrower and peaked at smaller separations than for a Sun-like sample. This is consistent with a steady decrease in average semi-major axis from the highest-mass binary stars to the brown dwarf binaries.
\end{abstract}

\keywords{binaries: general --- techniques: high angular resolution --- stars: late-type}

\section{Introduction}
\label{s:intro}

The multiplicity properties of stars hold clues to their formation and early evolution \citep[e.g.][]{goodwin2005,marks2011,bate2012}, and binarity is of fundamental importance for a range of astrophysical applications, such as determination of physical properties and target selection for exoplanet studies. Consequently, detailed multiplicity studies have been performed over a wide range of stellar masses and ages \citep[see e.g.][for recent summaries]{duchene2013,reipurth2014}. While multiplicity at the low-mass end -- in the M-dwarf regime -- has been a subject of study for a long time \citep[e.g.][]{fischer1992,delfosse2004,law2008}, there have recently emerged reasons to revisit this subject. The main reason for this is that the nearby M-dwarf population is becoming increasingly well characterized. Recent studies have greatly increased our sample of securely identified M-dwarf stars in the Solar neighborhood \citep[e.g.][]{riaz2006,reid2007,lepine2011}. Furthermore, while distances for this class of objects have previously been scarce due to the fact that they are generally too faint to have been observed by Hipparcos \citep{perryman1997}, recent parallax studies have started to become increasingly complete to the lowest-mass stars \citep[e.g.][]{henry2006,dittman2014,reidel2014}. Hence, larger well-defined statistical samples can be studied than has been possible before, and a greater accuracy is achievable in the characterization of their properties.

The AstraLux Norte \citep{hormuth2007a,hormuth2008} and Sur \citep{hippler2009} cameras are well suited for multiplicity studies by use of high-resolution imaging \citep[e.g.][]{hormuth2007b,daemgen2009,peter2012,bergfors2013}, with a resolving power of approximately 100~mas. AstraLux is a high speed and low read noise camera used for the purpose of so-called Lucky Imaging \citep[e.g.][]{tubbs2002,law2006}. Previously, we have used this instrument for the study of multiplicity in primarily early-type M-dwarfs \citep{bergfors2010,janson2012}. In the summary study of 2012 \citep{janson2012}, we found that the multiplicity properties of these stars were largely consistent with being continuously intermediate between the Sun-like \citep{raghavan2010} and brown dwarf \citep{burgasser2007} populations, though possibly with the exception of the mass ratio distribution \citep[see also][]{reggiani2013,goodwin2013}. The apparent continuities and discontinuities motivate further study of a later-type sample, bridging the gap between early/mid M-dwarfs in \citet{janson2012} and very low-mass (VLM) stars and brown dwarfs in \citet{burgasser2007}. The sample presented in \citet{lepine2011} provides an excellent basis for this purpose. Here we will present a study of multiplicity in mid/late M-type stars (primarily M3 and later, down to M8), which overlaps with both the previous M-dwarf and VLM studies.

In the following, we will first discuss the sample properties in Sect. \ref{s:sample}, and then the observations and data reductions in Sect. \ref{s:obs}. This will be followed by a summary of the results in Sect. \ref{s:results} and a description of the statistical properties of the sample in Sect. \ref{s:stats}. Finally, we will discuss the implications of this study in the context of multiplicity across all stellar masses in Sect. \ref{s:discussion} and summarize the conclusions in Sect. \ref{s:conclusions}.

\section{Target sample}
\label{s:sample}

\subsection{Observational properties}

The targets in this study were selected from the \citep{lepine2011} sample, where stars with a spectral type (SpT) estimate of M5 or later were selected if they were sufficiently bright ($J \leq10.0$~mag) and sufficiently far North ($>$-15$^{\rm o}$) to be meaningfully observed with AstraLux Norte. In total, this gave an input sample of 408 potential targets, of which 286 were actually observed. Targets from the `master list' of 408 stars were chosen entirely on the basis of observability during a given run and limited by the total amount of telescope time available for the program, hence the sub-selection of 286 actual targets can be seen as random, and should not introduce any selection effects in the analysis. The full set of observed targets is summarized in Table \ref{t:general}, where the basic observable quantities are from \citet{lepine2011} unless otherwise stated. In \citet{lepine2011}, the SpT estimates were not spectroscopically determined, but merely inferred from the $V-J$ colors of the stars. For our study, we have cross-matched these SpT estimates with actual SpTs in the literature for all cases where such measurements exist, and found that the former estimates exhibit a systematic offset toward later spectral types. For 198 out of the 286 observed stars, literature SpT determinations exist. Among these 198 cases, the median difference between the two estimates is 1 spectral sub-type. While a few extreme cases exist, such as I04122+6443 which is classified as M5 in \citet{lepine2011} but M1 in \citet{bender2008}, most stars are close to this 1 spectral sub-type offset. In Table \ref{t:general}, we adopt the literature SpT measurement for the 198 cases for which this is available, and denote the SpT with an upper case letter (e.g. `M5'). For the remaining 88 cases, we use the photometric estimations but label them with a lower case letter (e.g. `m5'), following the source notation. By analogy with the 198 overlapping cases, it is likely that the actual spectral type is approximately 1 spectral sub-type earlier than what the lower case notation implies for these 88 targets.

In 176 cases, we have been able to acquire trigonometric parallaxes. These have been  provided from a range of studies, the references for which are summarized in Table \ref{t:general}. Photometric parallaxes were used in the remaining cases, as provided in \citet{lepine2011}. Distances for the bulk of the sample range from 3 to 36~pc, with three targets at larger distances (40--70~pc). The median distance for the full sample is 15~pc. The 62\% coverage (176 out of 286) of trigonometric parallaxes is a substantial improvement on previous M-star studies such as \citet{janson2012}, in which the vast majority of distances had to be estimated photometrically. 

\subsection{Physical properties}

The fact that such a large fraction of the sample has trigonometric parallaxes is greatly beneficial for the estimation of semi-major axis distributions, as will be seen in Sect. \ref{s:separation}. On the other hand, the estimation of mass ratio distributions is very challenging for this class of objects. For late M-type stars, a mass cannot be reliably derived from spectral type alone, since the temperature of the object varies significantly during its long-lasting pre-main sequence phase\footnote{The pre-main sequence lasts about 180~Myr for a 0.2~$M_{\rm sun}$ star, 500~Myr for a 0.1~$M_{\rm sun}$ star, and 3~Gyr for a 0.075~$M_{\rm sun}$ star \citep[e.g.][]{burrows1993,burrows1997,baraffe1998}.}. Instead, masses have to be inferred based on models with significant uncertainties. These models also require the age of the system as an additional parameter, which is itself also highly uncertain in most cases. A combination of evolutionary and atmospheric models are required to make predictions for photometric values in a given band for a given stellar mass at a given age. Here we use both the NextGen \citep{hauschildt1999} and the more recent BT-Settl \citep{allard2014} atmospheric models, and the evolutionary models of \citet{baraffe1998,baraffe2003}. The COND \citep{allard2001} models are used to fill in some extreme ranges of the parameter space not covered by the aforementioned models. Differences between different evolutionary models are small compared to the other uncertainties considered here \citep[a few percent in luminosity for a given mass and age, see e.g.][]{burrows1997,saumon2008}.

Upper and lower boundaries for the ages are estimated in the following way: If a certain target has been identified as a member of a young moving group in the literature, the age boundaries of the moving group are assigned to the target irrespective of any other characteristics. These are estimated as 10--20~Myr for the $\beta$~Pic moving group \citep[e.g.][]{zuckerman2001,binks2014} and 50--150~Myr for the AB~Dor moving group \citep[e.g.][]{luhman2005a,janson2007}. Likewise, if a target is not identified as a moving group member but it has been subjected to a detailed age analysis in the literature, the corresponding age boundaries are assigned. For all other targets, we apply a rough age estimate based solely on their X-ray luminosity \citep[provided in][]{lepine2011}. If a target has a value of $L_{\rm X}/L_{\rm bol}$ comparable to the values of the targets studies in \citet{shkolnik2012}, it is assumed to have an age in the same range, and thus assigned 30~Myr as a lower bound and 300~Myr as an upper bound. If the value is lower but there is still detectable X-ray emission, the target is assumed to be older but still part of a young population with a lower bound of 300~Myr and an upper bound of 1~Gyr. If no X-ray emission is detected, it is assumed to be a field star with an age between 1~Gyr and 10~Gyr. The broad ranges are meant to encompass the fact that the uncertainties in the age determination are inevitably very large. Nonetheless, we strongly caution against taking the quoted age range for any individual target at face value; they should only be considered as broad general age assignments to the population, in order to benefit the statistical analysis.

\section{Observations and Data Reduction}
\label{s:obs}

All observations in this program were acquired with the AstraLux Norte camera on the 2.2m telescope at Calar Alto in Spain. The 2.2m telescope is on an equatorial mount. AstraLux uses an Andor DV887-UVB camera head equipped with a thinned, back-illuminated, electron-multiplying 512x512 pixel monolithic CCD. The CCD is equipped with two readout registers, one for conventional readout, and one 536 stage electron multiplication register. Each of the two registers comes with its own output amplifier. All Lucky Imaging data were obtained using the electron multiplication mode, and the associated output amplifier. The camera allows to select electron multiplication gains of up to 2500. For astronomical observations, the gain values are typically selected such that the ADU counts in the brightest pixel do not exceed 50\% of the linearity limit of the camera. It has been verified in lab experiments that charge transfer efficiency does not have any impact on the astrometric accuracy. Typical observations are made using a 256 by 256 pixel window readout, which facilitates shorter single frame integration times. The window also allows to avoid column 244 of the detector, which is subject to a charge trap that traps a few electron per clock cycle. Apart from column 244, the CCD has very good cosmetics without any clusters of bad pixels. The raw pixel scale (before oversampling; see below) is approximately 46~mas/pixel on average.

The observations were carried out in six separate runs: On 8--9 Nov 2011, on 5--8 Jan 2012, on 6--7 Jun 2012, on 27--29 Aug 2012, on 3 Sept 2012, and on 22--24 Nov 2012. Each target was observed in both the $i^{\prime}$-band and the $z^{\prime}$-band, and a large fraction of the targets, including the singles, were observed in two or more separate epochs. In total, excluding calibration observations, approximately 940 observations were acquired for the purpose of this survey, covering the 286 individual targets. As per usual, observing conditions varied during the runs, but since so many targets were observed several times, there is in general at least one frame of acceptable quality per star. The typical full width at half maximum (FWHM) is close to 100~mas, which is an appropriate measure for the resolving power of this instrument. For the purpose of astrometric calibration, we observed either the Trapezium or M15, depending on the season during which the observations were performed. The astrometric calibration is described in more detail in Sect. \ref{s:astro}.

Although the field of view of AstraLux Norte can be as large as 23\arcsec\ across with the full frame in use, many of the images were taken with a subarray readout and the target was not always centered perfectly in the frame, so the fully complete region has a radius of 5\arcsec around each star. We will thus only consider companions inside of 5\arcsec\ for statistical purposes in this study.

The basic data reduction makes use of the pipeline developed specifically for the purpose, described in \citet{hormuth2008}. The pipeline performs flat fielding and bias correction of the data, followed by a drizzle algorithm to oversample the image by a factor of two, for a final pixel scale of approximately 23 mas/pixel. Individual frames are then aligned based on the brightest pixel in the oversampled image, and the re-aligned images are subsequently recombined into collapsed images. By default, the pipeline produces four different reduced frames per full observation, corresponding to different cut-offs for the selection of frames used. In this study, we consistently used the selection in which the 10\% best frames were included in the collapsed frame. Individual frame exposure times were typically 30~ms, with minor variations depending on observing conditions and target brightness. The total number of frames was always selected so that the total integration time would add up to 300~s. Hence, the typical number of frames acquired was 10000, leading to a 10\% selection of 1000 frames, adding up to 30~s of `useful' integration time.

\section{Astrometry and photometry}
\label{s:astro}

Astrometry was first calculated in detector coordinates, and subsequently translated into sky coordinates using the calibration observations of the Trapezium or M15. For the calibration data, we chose five of the brightest stars in the field, determined their relative positions using Gaussian centroiding, and compared to the relative locations of these stars  in \citet{marel2002} for M15 and \citet{mccaughrean1994} for Trapezium. We also compared the results with a calibration based on the IRAF \textit{geomap} procedure \citep[see e.g.][]{kohler2008}, and found that calibration within a given observing run was consistent to within 1\% in pixel scale and 0.3$^{\rm o}$ in position angle, regardless of choice of calibration method and selection of stars within the method. In this way, it was found that the pixel scale and orientation of true North in the respective runs were: 23.57~mas/pixel and 1.66$^{\rm o}$ in Nov 2011, 23.58~mas/pixel and 1.72$^{\rm o}$ in Jan 2012, 23.67~mas/pixel and 1.90$^{\rm o}$ in Jun 2012, 22.59~mas/pixel and 1.83$^{\rm o}$ in Aug 2012, 22.69~mas/pixel and 1.96$^{\rm o}$ in Sept 2012, and 22.67~mas/pixel and 1.85$^{\rm o}$ in Nov 2012. The calibration errors are dominated by the 1\% uncertainty in pixel scale and the 0.3$^{\rm o}$ uncertainty in position angle mentioned above, which we adopt as the formal error bars in each case.

As in previous runs, astrometry for wide binaries in the sample was determined using Gaussian centroiding, and astrometry for close binaries was determined using PSF fitting \citep{bergfors2010}. Three PSFs were used in each case, to provide a well-defined mean and scatter in the PSF fitting. The PSFs were chosen among single stars in the survey to represent a broad range in observing conditions. In principle, one might tailor the PSF templates to each given target, such that only PSFs acquired under similar conditions are used in the fitting scheme. However, given the complex multi-modal variations of the PSF and the rapidly varying conditions during the observing nights, this is impractical, and our experience implies that no significant gain is achieved through such a procedure. Even if an apparent improvement were achieved, it would also be dubious whether the resulting implied precision could be trusted, given the aforementioned PSF complexity. We therefore consider it a better strategy to reflect a representative range of instrumental PSF realizations in the fitting, so that the derived error bars robustly encompass these variations. While the FWHM does not change much between different PSFs, since this measure is dominated by the diffraction-limited PSF core, beyond the core there can be quite a bit of variation in the PSF, sometimes showing diffraction rings and other times just a smooth halo. Astrometric values for the various systems are provided in Table \ref{t:astrometry}. Relative photometry was determined simultaneously with the astrometry, by measuring aperture photometry in the case of wide binaries, and the relative brightness of the PSFs fit to each component for close binaries. In the case of the so-called `false triple' effect, which often occurs in Lucky Imaging shift-and-add analysis when a binary with components of about equal brightness is observed and produces a tertiary ghost feature (at the same separation from the primary as the secondary but on the opposite side), we fit for all three components in the PSF fitting procedure. The photometry of the individual components was then calculated in the same way as in \citet{janson2012}, as first implemented by \citet{law2006thesis}.

With the age estimates from Sect. \ref{s:sample} in hand and the photometry derived here, upper and lower bounds for the individual component masses in each candidate multiple system are determined in the following way: Given a certain age estimate (an upper or lower bound for  given target), a grid of model values for each of $\Delta i^{\prime}$, $\Delta z^{\prime}$, and total $M_{\rm J}$ are calculated for every combination of possible primary and secondary masses covered by the parameter space of the theoretical atmospheric and evolutionary models\footnote{Photometric values in the relevant bands are provided directly by the models, such that no additional conversions are necessary.}. These model values are then compared to the actual measured values (with their measured error bars) for every real binary pair. The matching that provides the minimum $\chi^2$ then determines the masses that are assigned to the pair. For each star, we generate four different mass estimates, and for each binary pair we separately generate four different mass ratio estimates. The four estimates correspond to all possible combinations of the two age extremes and the two model sets \citep[NextGen and BT-Settl with associated evolutionary models, see][]{hauschildt1999,baraffe1998,baraffe2003,allard2014}. The final values and errors are then taken as the mean and the standard deviation of these four values for each star and each pair. In this way, both age and model uncertainties are considered in the estimations. The age uncertainty is the dominant one, due to the wide adopted ranges in this quantity. It is important to note that while the uncertainties in the individual stellar masses are large, the uncertainties in their mass ratios ($q = m_{\rm B}/m_{\rm A}$) are substantially smaller. This is due to the fact that any error in the age or model affects the estimated mass of the primary and secondary in a very similar way. As a result, while the median uncertainties in the primary and secondary masses are 23\% and 22\% respectively, the median error in the mass ratio is only 8\%. Table \ref{t:photometry} includes the masses and mass ratios that have been derived with this procedure.

\section{Detections and confirmations}
\label{s:results}

\begin{figure*}[p]
\centering
\includegraphics[width=16cm]{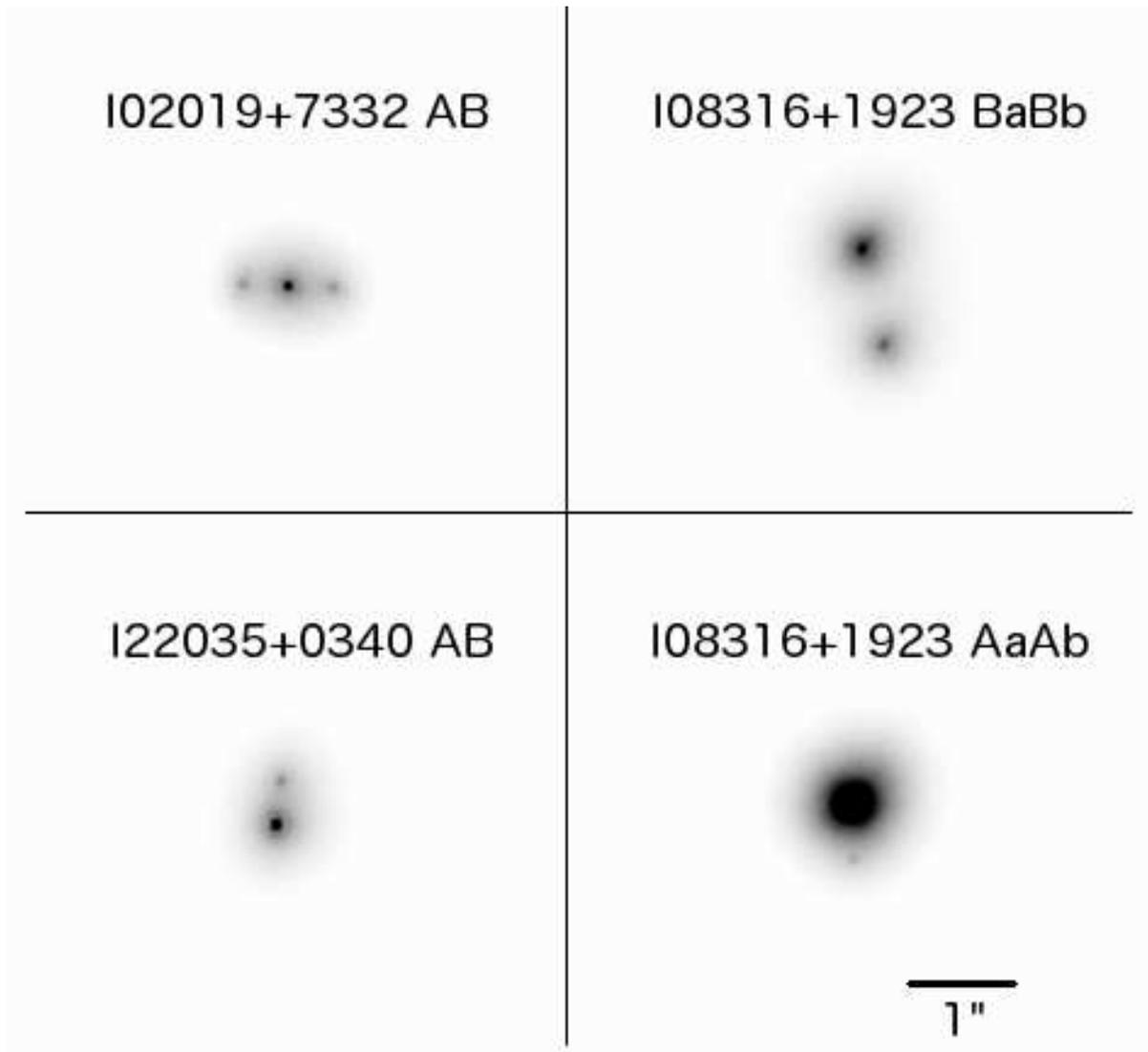}
\caption{Examples of multiples discovered with AstraLux in this survey. Top left: A close binary displaying the false triple effect that is common in such systems. Bottom left: A close binary without false triple effects. Top right: The Northern pair of the quintuple system I08316+1923, also known as GJ 2069. Bottom right: The Southern pair (with an additional unresolved companion to the Aa component) of the same quintuple system. The component farthest to the South marks a limiting case for what can be achieved with AstraLux Norte at this small separation. North is up and East is to the left in all images.}
\label{f:example_im}
\end{figure*}

Several both new and previously known companion candidates were detected in this survey, and many of them could be confirmed to share a common proper motion with the primary, confirming physical companionship. In total, 66 of the 286 systems were found to be either probable or confirmed multiples within the complete range of 5\arcsec separation, 41 of which were new discoveries. Of all systems, most were binaries and only two were triple systems, one of which was previously known. However, as noted in the individual notes, some of the systems are higher-order multiples when considering known companions outside of the AstraLux detectability range. Indeed, the system I08316+1932 is in reality a quintuple system, which is described in more detail in the individual notes. Several of the companions are probable brown dwarfs. A few examples of detected multiples are shown in Fig. \ref{f:example_im}.

For the candidates that were either observed twice with AstraLux or were already reported in previous imaging surveys, it was possible to test for common proper motion. Since these targets are very nearby and therefore have large proper motions in general, such a determination is possible even over rather short baselines. Our test followed the same structure as in \citet{janson2012} -- based on the location of a given candidate in one epoch relative to the primary star (in terms of separation and position angle), we made a prediction based on its proper motion and parallax of where it would occur in the second epoch if it were a static background object, and compared it to the actual measured position in the second epoch. If the locations were more than 3$\sigma$ discrepant, common proper motion was considered as confirmed. For candidates that passed this test, we also made a test for measureable orbital motion by testing if the first and second epoch positions differed from each other by more than 3$\sigma$. If so, we considered orbital motion as confirmed as well. These evaluations were based on the motion between the first and last listed data points for each given target listed in Table \ref{t:astrometry}, since this maximizes the observational baseline. After applying both tests, 37 candidates could be confirmed as bona fide companions, of which 33 also showed significant orbital motion. Three candidates could be discarded as background objects. 

A color test was applied to all 37 single-epoch candidates, in which it was checked whether the $\Delta i^{\prime}$ and $\Delta z^{\prime}$ yielded consistent results for an expected secondary. The same test was applied to one candidate for which two epochs of data exist, but where the baseline is insufficient for a conclusive proper motion test to be made. In this way, 33 candidates were found to have colors consistent with real companions. Five candidates were too blue to be low-mass stellar companions ($\Delta z^{\prime} - \Delta i^{\prime} > 0$, which would imply that the secondary is bluer than the primary), and thus discarded as likely background contaminants, although astrometric follow-up in the future will still be valuable for such candidates, in order to test whether there could be white dwarf companions among them. Even for many of the candidates that have only been observed or detected in one epoch, it is possible to draw conclusions about common proper motion. The targets move rapidly across the sky (from $\sim$100~mas/yr to several hundreds of mas/yr), and have been observed in previous all-sky surveys spanning decades backward in time. Hence, any background contaminant that happens to end up close to the primary star at the AstraLux epoch should be separated from it by up to several arcseconds in those previous epochs of data. Hence, they are often detectable there, despite the much worse spatial resolution of wide-field surveys, and so from their presence or absence in the archival data, it can be determined whether or not they share a common proper motion with the primary. We have used archival data from primarily two surveys for this purpose: The Two Micron All Sky Survey \citep[2MASS, see][]{skrutskie2006} and the first Palomar Observatory Sky Survey (POSS). Since 2MASS was performed in the late nineties up across the milennial shift, it provides up to a 15 year baseline, and a quite reasonable spatial resolution for a wide-field survey. However, while POSS has a slightly worse spatial resolution, it is the most useful survey for this purpose. This is due to the fact that it was performed largely in the early 1950s, providing a 60 year baseline for the vast majority of the targets. Since the candidates are bright, sensitivity is not a limiting issue for these purposes, but the most important issue is how far a backgrund contaminant would have traveled relative to the primary since the archival epoch, hence why a large baseline is preferred. By examining these archival data sets, we were able to conclude for 24 targets that if the candidate were a background contaminant, it would have been clearly visible in the images. Since they are not there, we can infer that the candidates are physical companions that share a common proper motion with the primary. In most of the 9 remaining cases (which are generally the targets that have the slowest proper motions and/or the faintest companions), a background contaminant would have been marginally detectable, but for any such limit case, we count common proper motion as not having been proven yet. 

The vast majority (and probably all) of these 9 remaining cases are expected to be real companions. Aside from the high confirmation rate in the candidates for which a proper motion test has been performed, this can also be deduced from the fact that the distribution of the candidates in projected separation is strongly slanted toward small separations, while the opposite would be true in a sample dominated by background contaminants. They also all pass the color test mentioned above, matching the expectation for physical companions, which would be rare for background contaminants, since the blackbody flux peak sweeps across the $i^{\prime}-z^{\prime}$ wavelength range in the M-dwarf regime. In total, we thus consider 68 candidates in 66 systems to be either probable or confirmed physical companions.

\begin{figure}[p]
\centering
\includegraphics[width=8cm]{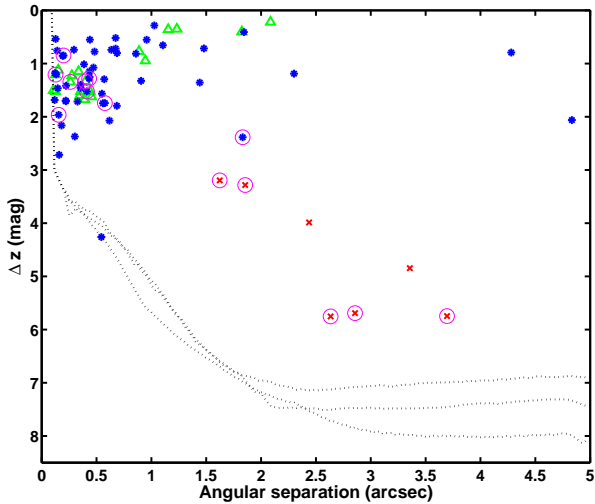}
\caption{Plot of the AstraLux detections in angular separation versus $\Delta z^{\prime}$. Red crosses are confirmed or suspected background stars. Green triangles are confirmed or probable binaries that are estimated as having been positively selected for (i.e., that would have been too faint to make the selection cut if the primary had been single). The blue asterisks are the `statistically clean' (see Sect. \ref{s:multfrac}) confirmed or probable binaries. Pairs for which either physical companionship or background contamination is probable but has not yet been demonstrated through common proper motion are encircled in magenta. Also plotted are the median contrast curves for the faint (top), intermediate (middle) and bright (bottom) targets (see text).}
\label{f:sep_vs_dz}
\end{figure}

The detections are plotted in Fig. \ref{f:sep_vs_dz}, and the binary properties are summarized in tables \ref{t:astrometry} and \ref{t:photometry}. 

\section{Statistical analysis}
\label{s:stats}

\subsection{Multiplicity fraction}
\label{s:multfrac}

In order to translate the 66/286 multiple systems into an actual multiplicity fraction, we need to take a number of subtle bias and selection effects into account. One of the most important factors in this regard is the brightness-limited nature of the sample. We imposed a constraint of 10~mag in $J$-band when selecting the targets. This will cause an excess of binaries in the sample, because for some binaries, the primary will be fainter than 10~mag, but the sum of the primary plus secondary light will be brighter than this limit. Hence, these binaries will be selected into the sample only because they are binaries, and would not have been selected if they were single. We can account for this effect by identifying those binaries that have been positively selected for, and simply removing them from the sample for the purpose of calculating a multiplicity fraction\footnote{We also note that if this correction is not done, there will be an artificial strong peak toward near-equal masses in the mass ratio distribution, as we demonstrated in \citet{janson2012}}. This is done by calculating individual $J$-band magnitudes for each component of each multiple system, using the measured $\Delta z^{\prime}$ as a proxy for the $\Delta J$ value, and discarding those cases for which the primary $J$-band magnitude becomes fainter than 10~mag. A total of 18 systems are found to have been positively selected for, which leaves a sample in which 48 out of 268 systems are multiple (referred to henceforth as the `statistically cleaned' sample). This results in a multiplicity fraction inside of the AstraLux sensitivity range of $48/268 = 17.9$\%. 

In order to estimate a total multiplicity fraction that is independent of the AstraLux sensitivity, an assumption of the underlying distributions in mass ratio and semi-major axis needs to be made, and the corresponding population needs to be related to the AstraLux sensitivity space in order to evaluate what fraction of binaries fall into this space (the `detectable fraction') and which fraction does not. As we will see, this is a complicated issue for this type of sample, where the mass ratio distribution is unconstrained for small mass ratios. Given the range of distributions that fit the data as discussed in Sect. \ref{s:massratio}, the detectable fraction probably lies between 66.6\% and 85.4\%. This gives a range of possible multiplicity fractions from $48/268/0.854 = 21.0$\% to $48/268/0.666 = 26.9$\%. Hence, the uncertainty on the multiplicity fraction arising from the unknown mass ratio distribution is comparable to the random (Poisson distributed) error, which is approximately $\pm$3\%.

\subsection{Semi-major axis distribution}
\label{s:separation}

Given that a significant fraction of the stars in our sample have trigonometric parallaxes, we can establish good projected physical separations in general, which benefits the purpose of determining a well-constrained semi-major axis distribution. For translating between projected physical separation and semi-major axis, we use the same conversion factor of close to 1 as in \citet{janson2013}, based on the derivation of \citet{brandeker2006} for a typical eccentricity distribution of $f(e) \sim 2e$. Our procedure for determining the semi-major axis distribution is based on generating a simulated population with a certain distribution, subjecting it to the sensitivity limits of AstraLux, and testing how well the resulting sample matches the actual body of detections, using a Kolmogorov-Smirnov test. 

\begin{figure}[p]
\centering
\includegraphics[width=8cm]{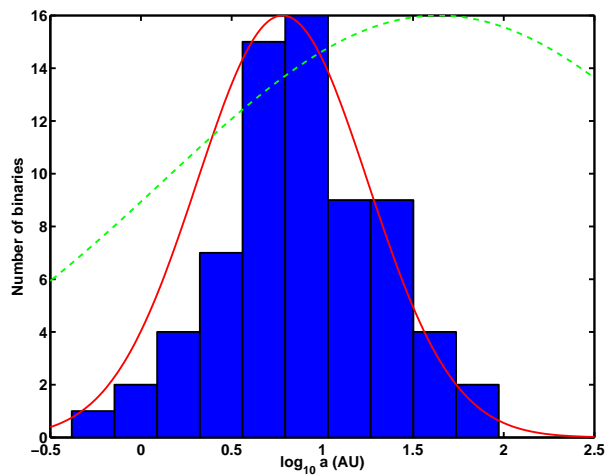}
\caption{Distribution in semi-major axis of the sample. The histograms are the estimated semi-major axes of the observed sample. The red curve is the best-fit distribution in our simulations. Note that the simulations take into account the incompleteness effects at small separations, hence why the measured distribution sits lower than the model distribution at small separations. The green dashed curve is the corresponding distribution for Sun-like stars \citep{raghavan2010}, which is clearly too broad and peaked at too large values to match this late M-dwarf sample.}
\label{f:semimaj_distr}
\end{figure}

To begin with, we will assume that the sample has a uniform mass distribution, but later on we will discuss how changing the mass ratio distribution affects these results. As input distributions for the semi-major axes (in units of AU here), we choose log-normal functions, both since this is the usual choice in this type of study \citep[e.g.][]{duquennoy1991,raghavan2010,janson2012} and since it a priori appears to potentially provide a good fit to the observed distribution (see Fig. \ref{f:semimaj_distr}). We then vary the $\sigma_{\rm a}$ and $\mu_{\rm a}$ parameters of the distribution in steps of 0.01 and see how the choices in these parameters affect the quality of the fit to the observed distribution. The steps are performed in a grid where both the $\sigma_{\rm a}$ and $\mu_{\rm a}$ values are varied simultaneously, in order to find the global maximum in fit quality. This is important since there is some covariance in these parameters, where (e.g.) a smaller $\sigma_{\rm a}$ can potentially be partly compensated for through a larger $\mu_{\rm a}$, and vice versa.

The outer boundary of the AstraLux sensitivity range is set by the 5\arcsec\ completeness radius. The inner boundary is set by the resolving power of approximately 100~mas. In between, the detectability of a candidate is set by the brightness contrast of the candidate relative to the contrast curve of the instrument. Contrast curves are calculated in the same way as is typically done in imaging surveys for faint companions \citep[e.g.][]{lafreniere2007,janson2011}, by taking the standard deviation in a series of annuli centered on the star with different radii to represent the $\sigma$ at various separations from the star, and relating them to the measured flux of the star for representing the limiting contrast of detectability. A 5$\sigma$ criterion is chosen as the basis for the contrast curves. Since the resulting contrast curve varies a bit with the brightness of the primary, we have divided the target stars into faint (9--10~mag), intermediate (8--9~mag) and bright ($<$8 mag). Representative contrast curves are then derived by taking the median of the contrast curves for all single stars in the survey in each brightness category. The simulated populations are set to have the same brightness distribution as the full real sample, and so the detectability of companions around a given simulated star is evaluated based on the representative contrast curve for its particular brightness category. Any companion in the simulation that ends up inside of these completeness boundaries is counted as being detected, and any companion that does not is counted as a non-detection.

\begin{figure}[p]
\centering
\includegraphics[width=8cm]{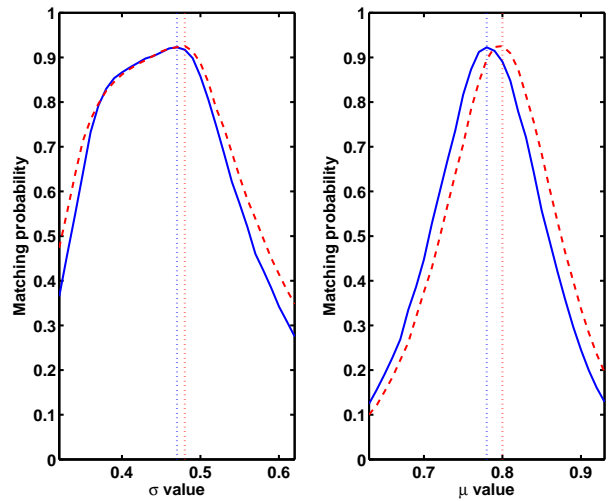}
\caption{Match probabilities from the test in which the $\mu_{\rm a}$ and $\sigma_{\rm a}$ parameters of the simulated Gaussian semi-major axis distribution function were varied, and the resulting distribution tested against the measured sample. Blue solid lines represent simulations based on an underlying uniform mass ratio distribution, and red dashed lines represent simulations based on a linearly increasing mass ratio distribution. Left: The distribution of probabilities as function of $\sigma_{\rm a}$, for a cross-section in the parameter grid along $\mu_{\rm a} = 0.78$ in the uniform case, and along $\mu_{\rm a} = 0.80$ in the linearly increasing case. Right: The distribution of probabilities as function of $\mu_{\rm a}$, for a cross-section in the grid along $\sigma_{\rm a} = 0.47$ in the uniform case, and along $\mu_{\rm a} = 0.48$ in the linearly increasing case. Dotted lines denote the location of the global probability maximum in each case. As can be seen, there is a well-defined maximum in each distribution, and the uniform and linearly increasing cases give very similar results, showing that the underlying mass ratio distribution does not significantly affect the determination of the semi-major axis distribution.}
\label{f:traces}
\end{figure}

Finally, the separation distribution of `detected' simulated companions is compared to the distribution of the actual detected sample. Every test is done 1000 times, and the median of the match probability of the 1000 tests is adopted \citep{babu2006}. For the log-normal distribution of the semi-major axis $a$ in units of AU, we find that $\mu_{\rm a} = 0.78$ and $\sigma_{\rm a} = 0.47$ gives the best match to the observed distribution, with a match probability of 92.4\%. As we mentioned previously, this is under the assumption of a uniform mass ratio distribution. If we instead choose a linearly increasing mass ratio distribution, the best-fit values become $\mu_{\rm a} = 0.80$ and $\sigma_{\rm a} = 0.48$ with a match probability of 92.5\%. Hence, the result is not heavily dependent on the mass ratio distribution. The match probabilities as function of $\mu_{\rm a}$ and $\sigma_{\rm a}$ for cross-sections in the parameter grid with values of $\pm$0.15 around the best-fit values are shown in Fig. \ref{f:traces}, for both the cases of a uniform and a linearly increasing underlying mass ratio distribution.

\section{Mass ratio distribution}
\label{s:massratio}

\begin{figure}[p]
\centering
\includegraphics[width=8cm]{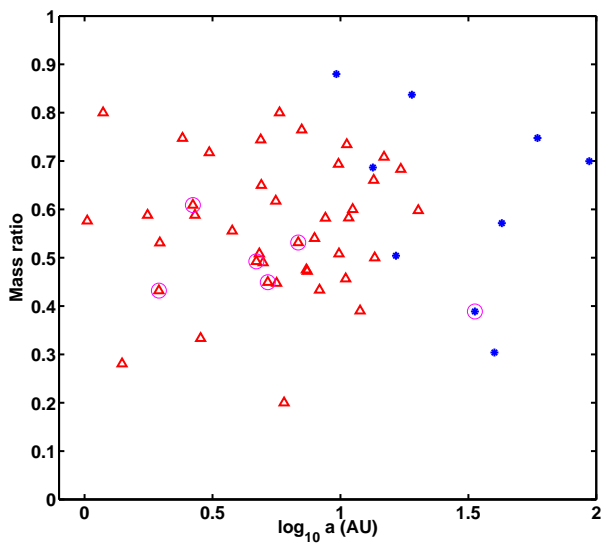}
\caption{Estimated semi-major axis versus mass ratio for the statistically clean sample (see Sect. \ref{s:multfrac}). The red triangles are inside of 1\arcsec\ projected separation, which makes them unsuitable for assessing a mass ratio distribution due to the false triple bias discussed in the text. Hence, only the blue asterisks are used for this purpose. Binaries for which physical companionship is probable but has not yet been demonstrated through common proper motion are encircled in magenta.}
\label{f:aest_vs_mfrac}
\end{figure}

As has been mentioned previously, the mass ratio distribution is very challenging to constrain. This is due to several reasons: 1) It is difficult to assign reliable masses to late M-type stars due to uncertainties in age and evolutionary models, although as we have seen in Sect. \ref{s:sample}, this has a relatively small impact on the mass ratio. 2) The survey is incomplete for the smallest mass ratios, meaning that the distribution cannot be well constrained there. 3) There appears to be a bias avoiding near-equal brightnesses in close systems that are subject to the false triplet effect \citep[see][]{janson2012}. This affects the mass ratio distribution in a way that is difficult to quantify. In order to mitigate the third issue, we only consider binaries outside of 1\arcsec\ separation (see Fig. \ref{f:aest_vs_mfrac}). This completely avoids the bias, but also leaves us with a smaller sample, such that less stringent conclusions can be drawn about the mass ratio distribution than the semi-major axis distribution.

\begin{figure}[p]
\centering
\includegraphics[width=8cm]{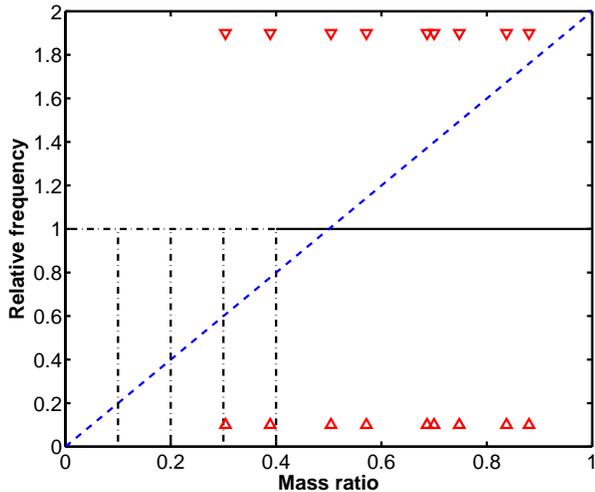}
\caption{Illustration of the various mass ratio distributions used in the simulations. The black lines represent a uniform distribution in mass ratio. The dash-dotted lines represent different possible choices of a lower cut-off in mass ratio $q_{\rm min}$. The dashed blue line represents a linearly increasing mass ratio distribution. Each pair of red arrows represents the mass ratio of a binary pair used in the test (most detected pairs in the survey have $<$1\arcsec separations, and are therefore excluded in this analysis). The mass ratio distribution is only loosely constrained, and all of the distributions illustrated here are formally consistent with the observational data.}
\label{f:only_mfrac}
\end{figure}

The procedure for determining the distribution is the same as in the previous section, apart from that we consider 1\arcsec\ instead of 100~mas as the effective inner working angle, and obviously that we compare the mass ratio distributions of the real and simulated samples instead of the semi-major axis distribution. We test three cases of mass ratio distributions: a uniform distribution ($f \sim q^0$), a linearly increasing distribution ($f \sim q^1$), and a uniform distribution but with a cut-off at some minimum mass ratio $q_{\rm min}$, for which we test a range of values. These different cases are illustrated in Fig. \ref{f:only_mfrac}. As a semi-major axis distribution, we simply choose  $\mu_{\rm a} = 0.78$ and $\sigma_{\rm a} = 0.47$ here -- as with the reverse case, the specific choice of semi-major axis distribution does not affect the results in any significant way. The uncertaintes in the measured mass ratios are represented by assigning Gaussian distributed random errors given by the estimated values listed in Table \ref{t:photometry} to the mean mass ratios, with a different random seed for each simulation. 

We find that a uniform distribution provides a match probability of 58.0\%, which is an entirely reasonable fit to the data. However, the distribution is unconstrained at small mass ratios. For instance, a uniform distribution with a cut-off at $q_{\rm min} = 0.3$ gives a match probability 84.1\%, which is an even better fit. We therefore step through $q_{\rm min}$ in steps of 0.01 in order to test when it becomes marginally inconsistent with the data, which we count here as a match probability of less than $\sim$33\%, i.e. equivalent to a typical 1$\sigma$ rejection. This occurs at a $q_{\rm min}$ of 0.39. A linearly increasing mass ratio gives a match probability of 53.0\%. This is almost equally consistent with the data as the fully uniform distribution, underlining the fact that the mass ratio distribution is largely unconstrained.

The range of $q_{\rm min}$ that fit the data ($\sim$0.0--0.39) give a range of possible detectable fractions, which feeds back into the multiplicity fraction discussed in Sect. \ref{s:multfrac}. The lowest $q_{\rm min}$ (fully uniform distribution) gives a detectable fraction of 66.6\%, and the highest gives a fraction of 85.4\%.

\section{Discussion}
\label{s:discussion}

\begin{figure}[p]
\centering
\includegraphics[width=8cm]{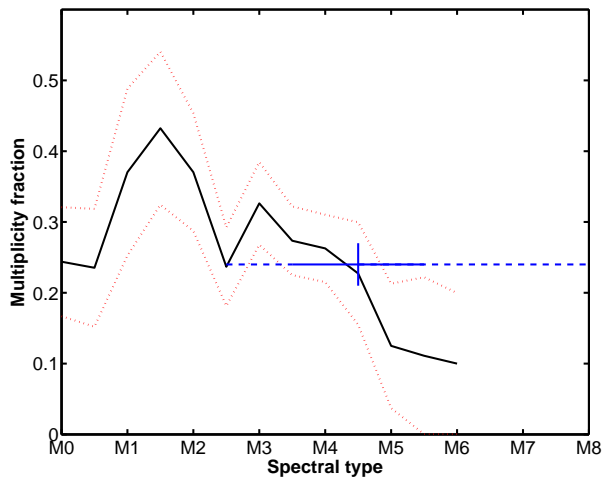}
\caption{Multiplicity fraction of low-mass stars as function of spectral type. The solid black and dotted red lines are multiplicity values (mean and error bars) from \citet{janson2012}. The blue orthogonal lines represent the multiplicity in this sample. The solid horizonthal line is the median (M4.5) plus/minus one standard deviation (1 spectral sub-type) for the spectral types in the sample. The dashed line represents the full spectral type range of the sample, excluding I04122+6443 which was classified as M5 in \citet{lepine2011} but M1 in \citet{bender2008}. The multiplicity fraction derived here is consistent with the results from \citet{janson2012}.}
\label{f:multfrac}
\end{figure}

As we have seen, the mass ratio distribution is one of the main contributors to uncertainty in the total multiplicity fraction. Aside from the caveats already mentioned in the mass ratio determination, it could also be the case that the mass ratio distribution has a dependence on semi-major axis. Such a dependence has been hinted at in several other multiplicity studies \citep[e.g.][]{janson2013,lafreniere2014}. If so, the mass ratio distribution determined at $>$1\arcsec\ may not be representative for the $<$1\arcsec\ region where the majority of binaries reside. Nonetheless, it appears that values of $\sim$21--27\% appropriately bracket the most plausible total multiplicity fraction range. This range is fully consistent with a smoothly decreasing multiplicity fraction as function of primary mass, as arrived at in many previous studies \citep[e.g.][]{kouwenhoven2007,raghavan2010,janson2012}. In Fig. \ref{f:multfrac}, we show a comparison between the multiplicity fraction of our full sample and the multiplicity as a function of spectral type from \citet{janson2012}. Our derived multiplicity is well consistent with this previous study. The two studies imply a smooth evolution across the M-type range, with no evidence for any sudden jumps, which has been suggested in some scenarios that consider star and brown dwarf formation as separate processes \citep[e.g.][]{thies2007}.

It is currently not possible to distinguish stringently whether the mass ratio distribution remains close to uniform toward small mass ratios, or whether it starts to decrease somewhere below $q = 0.4$. It is in any case clear that there is no sharp cut-off below a $q$ of $\sim$0.8, as has been reported for the yet lower-mass sample of VLM stars and brown dwarfs \citep[e.g.][]{burgasser2007}. There could however in principle be such a cut-off in the $q < 0.4$ range, as our analysis with $q_{\rm min}$ in Sect. \ref{s:massratio} demonstrates. If so, the necessarily lower threshold of our sample could imply that there would be some characteristic secondary mass for which companions become less frequent. However, this should obviously be taken as mere speculation at this point, given the incompleteness issues, in addition to the difficulties in the mass ratio determinations.

The semi-major axis distribution, on the other hand, is well constrained by the data, since the AstraLux sensitivity range covers the majority of the range of where the binaries reside, encompassing both sides of a distribution that is well represented by a Gaussian function. As derived in Sect. \ref{s:separation} and shown in Fig. \ref{f:semimaj_distr}, a Gaussian distribution with $\mu_{\rm a} = 0.78$ and $\sigma_{\rm a} = 0.47$ matches the data at $>$90\% probability. By contrast, a Sun-like distribution with $\mu_{\rm a} = 1.64$ and $\sigma_{\rm a} = 1.52$ \citep{raghavan2010} only has a $<0.03$\% probability of matching the data and can be firmly excluded. Thus, the result fits the trend of a semi-major axis that gets continuously narrower and closer in with decreasing primary mass \citep[e.g.][]{burgasser2007,janson2012}, with an opposite trend toward higher masses \citep[e.g.][]{kouwenhoven2007,janson2013}.

Most trends observed here and in other studies of low-mass stars are consistent with a smooth transition from the highest-mass stars to the lowest mass brown dwarfs in the field \citep[e.g.][]{luhman2005b,bourke2006}, possibly implying a universal formation scenario for this whole range of objects. The main remaining mystery in this regard is the mass ratio distribution, which appears to be markedly different between stars and brown dwarfs \citep[e.g.][]{goodwin2013}. Given the many difficulties in assigning reliable masses to low-mass objects, however, it would be valuable to further study the mass ratios of both our systems and systems of yet lower mass. We are currently looking into ways in which this could be done. A high-resolution spectrograph with adaptive optics capacity, such as CRIRES at the VLT \citep{kaufl2004}, could measure the individual radial velocities of the components of binaries discovered here, and thus could directly measure model-independent mass ratios over a relatively short timeframe. This would greatly assist studies of multiplicity properties of low-mass stars and brown dwarfs, by aiding in both mass ratio distribution and multiplicity fraction determinations.

\section{Conclusions}
\label{s:conclusions}

We have presented observations of 286 mid/late M-dwarfs using the AstraLux Norte camera at Calar Alto in Spain. We resolved 66 probable or confirmed multiple systems, of which 41 were new discoveries. The majority of binary candidates were observed twice or more, and could be confirmed as bona fide companions.

Based on these discoveries and evaluations of the sensitivity range of AstraLux Norte, we deduced a multiplicity fraction inside of the AstraLux sensitivity range of 17.9\%, corresponding to a total multiplicity fraction of 21--27\%. The mass ratio distribution is consistent with being uniform down to $q = 0.4$, but cannot be stringently constrained below this value. The semi-major axis distribution is well represented by a Gaussian function with $\mu_{\rm a} = 0.78$ and $\sigma_{\rm a} = 0.47$ -- a function which is significantly narrower and peaked at smaller separations than the corresponding distribution of Sun-like stars.

Most observables point to continuous distributions and a common formation scenario for stars and brown dwarfs, but some discrepancies persist, most notably in the mass ratio distribution. This is however also one of the most uncertain distributions, and more work will be required in the future to more robustly assess mass ratios at the low-mass tail of the stellar population.

\acknowledgements
We thank all the staff at the Calar Alto observatory for their support. This study made use of the CDS services SIMBAD, VizieR, and CDS Portal, as well as the SAO/NASA ADS service. Some of the archival study was based on photographic data of the National Geographic Society -- Palomar Observatory Sky Survey (NGS-POSS) obtained using the Oschin Telescope on Palomar Mountain. The NGS-POSS was funded by a grant from the National Geographic Society to the California Institute of Technology. The plates were processed into the present compressed digital form with their permission. The Digitized Sky Survey was produced at the Space Telescope Science Institute under US Government grant NAG W-2166. Other parts of the archival study make use of data products from the Two Micron All Sky Survey, which is a joint project of the University of Massachusetts and the Infrared Processing and Analysis Center/California Institute of Technology, funded by the National Aeronautics and Space Administration and the National Science Foundation.

\appendix
\section{Notes on individual targets}
In this section, we list special remarks on individual targets, where relevant.

\textbf{I00235+7711 (GJ 1010)}
As noted in the Washington Double Star (WDS) catalogue \citep[e.g.][]{mason2001}, this star is a member of an 11\arcsec\ binary. Both components are visible in full-frame AstraLux images, but since the separation is larger than the region of completeness, I00235+7711 counts as a single star within this range.

\textbf{I00395+1454N (G 32-37 B)}
As implied by its identifier, I00395+1454N has a wide binary companion toward the south at a separation of 17\arcsec\ according to the WDS, which is not included in the AstraLux field of view. Given the new detection of a closer companion with AstraLux, it is probable that the system is in reality at least a triple system.

\textbf{I00413+5550W (GJ 1015 A)}
I00413+5550W has a 10\arcsec\ companion to the East that is noted in WDS. It is visible in the AstraLux image but outside of its completeness range, and thus the star counts as single for statistical purposes.

\textbf{I01028+4703 (G 172-35)}
This target is a component of a 15\arcsec\ binary noted in WDS, which is visible in the AstraLux images, but outside of the completeness range.

\textbf{I01076+2257E (GJ 9040 B)}
As its name implies, I01076+2257E is a secondary component in a 10\arcsec binary noted in WDS, which can be seen in the AstraLux images, but does not count in our analysis due to its $>$5\arcsec separation.

\textbf{I02027+1334 (GJ 3129)}
This target has been noted as a double-lined spectroscopic binary with an estimated maximum semi-major axis of 0.13~AU \citep{shkolnik2010}. As expected, it is therefore not detected in AstraLux imaging. There are no other companions seen with AstraLux, and it thus counts as single in the statistics.

\textbf{I03325+2843 (J03323578+2843554)}
This triple system has been observed in several epochs, as reported in \citet{janson2012}. We will discuss it in more detail in a near-future study (Janson et al., in prep.) that analyzes systems observed in multiple epochs with AstraLux. The BC pair doesn't count as a separate pair in the statistical analysis, since its separation is just below 100~mas in the epoch considered here. The measured brightness differences between the B and C components are $\Delta z^{\prime} = 0.58 \pm 0.12$~mag and $\Delta i^{\prime} = 1.03 \pm 0.08$~mag.

\textbf{I03372+6910 (GJ 3236)}
The I03372+6910 system is a known double-lined spectroscopic binary \citep{shkolnik2010}, which is also eclipsing \citep{irwin2009}, and thus is of significant use for calibration of low-mass stellar properties. The companion is at a $<$0.1~AU separation and thus beyond the AstraLux sensitivity range, where we find no additional companions.

\textbf{I03392+5632 (G 175-2)}
In addition to the companion discovered in this study, there is a known wide (24\arcsec) common proper motion companion noted in WDS, hence the system is at least triple in reality.

\textbf{I04123+1615 (LP 414-117)}
There is a spectroscopic binary companion to this star with an orbital period of $\sim$128~days \citep{bender2008}. It is too close to be spatially resolved with AstraLux, and there are no other companions within the sensitivity range of our study.

\textbf{I04129+5236 (LHS 1642)}
I04129+5236 is a known close binary system with a well-determined orbit \citep{pravdo2004,martinache2009}. Its separation is smaller than 100~mas at all times, and it therefore remains unresolved by AstraLux. We do detect one other point source in the field of view, but it is a suspected background contaminant based on its blue color, with $\Delta z^{\prime} = 5.8 \pm 0.1$~mag and $\Delta i^{\prime} = 5.3 \pm 0.1$~mag.

\textbf{I04247-0647 (J04244260-0647313)}
This is a target that overlaps with our previous study in \citet{janson2012}. As we already noted there, it is single in the AstraLux field of view, but it is a triple-lined spectroscopic multiple system further in \citet{shkolnik2010}.

\textbf{I04308-0849S (Koenigstuhl 2 B)}
Another target that overlaps with \citet{janson2012}, this star is single in the AstraLux field but has a known wide common proper motion companion at 17\arcsec\ separation \citep{caballero2007}.

\textbf{I04388+2147 (G 8-48)}
In addition to the companion discovered here, there is a wide binary companion at 15\arcsec\ separation noted in the WDS, hence it is likely that the system is at least a triple.

\textbf{I04425+2027 (J04423029+2027115)}
This is a double-lined spectroscopic binary with a period of a few days \citep{mochnacki2002}, which is far too close to be resolved with AstraLux. It is single in our sensitivity range.

\textbf{I05030+2122 (LP 359-186)}
For binaries with a small brightness difference between the components, one should always be wary of potential 180$^{\rm o}$ phase shifts between different astrometric epochs. For this system, such a shift in the \citet{law2008} data point with respect to our two epochs seems highly probable, given the consistency in apparent orbital motion when such a shift is considered ($\sim$1$^{\rm o}$ per year in each case with the shift included, versus a sudden change from $\sim$30$^{\rm o}$ per year to $\sim$1$^{\rm o}$ per year when it is not). The quoted value in Table \ref{t:astrometry} therefore includes such a shift.

\textbf{I06171+0507 (NLTT 16333)}
I06171+0507 is a close binary that has been previously resolved in several epochs by \citet{pravdo2006}, and which we re-detect with AstraLux. The pair is itself part of a higher-order multiple system with the bright star HR~2251 at a 103\arcsec\ separation.

\textbf{I06579+6219 (GJ 3417, LHS 1885)}
If taken at face value, the body of astrometric points that exists for this target does not make sense. Three epochs of astrometry exists: One from \citet{henry1997} taken in 1996 ($\rho = 2.0$\arcsec\ and $\theta = 220^{\rm o}$), one from \citet{law2008} taken in 2005 ($\rho = 1.5$\arcsec\ and $\theta = 320^{\rm o}$), and our data point taken in 2012 ($\rho = 1.4$\arcsec\ and $\theta = 240^{\rm o}$). Given that background objects can be firmly excluded, this would imply an enormously fast orbital motion since the binary would move from 220$^{\rm o}$ to 320$^{\rm o}$ and then back to 240$^{\rm o}$ within 16 years, which is impossible for such a low-mass binary with a $\sim$17~AU semi-major axis. However, the astometry becomes entirely sensible in an orbital motion framework if we impose a 90$^{\rm o}$ phase shift on the \citet{law2008} position angle such that it is 230$^{\rm o}$ instead, giving a continous motion of 20$^{\rm o}$ in 16 years. In \citet{janson2012}, we suggested an equal phase shift for similar reasons for the \citet{law2008} data point in the J15553178+3512028 binary system. We thus include such a shift in Table \ref{t:astrometry}. 

\textbf{I07111+4329 (J07111138+4329590)}
I07111+4329 is a known binary that has been observed previouly over several epochs \citep[e.g.][]{dupuy2010}. There is also a background star in the field of view, with $\Delta z^{\prime} = 5.7 \pm 0.1$~mag and $\Delta i^{\prime} = 4.2 \pm 0.1$~mag.

\textbf{I07307+4811 (LHS 229)}
Although this star looks single in the AstraLux images, it is in fact part of a quadruple system \citep{harrington1981}. I07307+4811 itself is a close binary M-dwarf pair with a separation of $\sim$50~mas, too close to be resolved here. In addition, at a 103\arcsec separation far beyond the AstraLux field of view, there is a pair of white dwarfs that are physically bound to this system.

\textbf{I07320+1719W (G 88-35)}
As implied by its identifier, I07320+1719W is part of a wide binary system registered in WDS with an 11\arcsec\ separation. It is visible in the AstraLux images, but beyond the completeness range of the instrument.

\textbf{I08119+0846 (LHS 35)}
There is a relatively strong linear trend noted in the radial velocity analysis for the target in \citet{bonfils2013}. Such trends can be signs of stellar companions, but in this case we detect no companions in the AstraLux data.

\textbf{I08316+1923 (CU Cnc and CV Cnc)}
This is a known quintuple system, as reported in e.g \citet{delfosse1999} and \citet{beuzit2004}. Four of the components are resolved in two separate pairs (AaAb and BaBb) by AstraLux. The two pairs themselves (CU~Cnc and CV~Cnc) are too far separated for the AstraLux completeness range, and so they count as two separate binary pairs for statistical purposes. The fifth component is unresolved in the images; this is an ecipsing binary companion to the Aa component.

\textbf{I08589+0828 (G 41-14)}
I08589+0828 is a triple system with a close spectroscopic pair on a 7.6 day orbit, and one wider component which is reported at a separation of 620~mas in \citet{delfosse1999}. Subsequently, an orbit has been determined for the wider pair \citep{hartkopf2012}, with a period of 5.66 years and a 424 mas angular semi-major axis. Using the estimated orbital elements to predict the location of the wider component in Jan 2012 when the AstraLux image was taken, the predicted separation is $\sim$100~mas. This is in excellent agreement with what is seen in the AstraLux image, where the PSF is substantially extended, but not quite sufficiently to get a satisfactory binary fit. Since the fit does not converge, the star counts as single within the AstraLux sensitivity range.

\textbf{I10497+3532 (GJ 1138)}
I10497+3532 has been previously reported as a 300~mas binary \citet{beuzit2004}. The fact that it looks single in our AstraLux images despite the excellent quality of those observations implies that it must have undergone substantial orbital motion, bringing it to a much smaller ($<$100 mas) projected separation in June of 2012. The non-detectability in AstraLux images means that it counts as a single system for the purpose of the multiplicity analysis performed here.

\textbf{I13143+1320 (NLTT 33370)}
Recently, \citet{schlieder2014} reported I13143+1320 as a binary. In NACO images that are approximately coincident with the AstraLux images, the projected separation of the binary is $\sim$75~mas, which is consistent with the fact that the binary is unresolved in the AstraLux images. It counts as a single system in the analysis performed here.

\textbf{I15474+4507 (G 179-55)}
This star is a known eclipsing and double-lined spectroscopic binary with a period of 3.55 days \citep{hartman2011}. The separation is far too small to be resolved with AstraLux, and we detect no other companions within the AstraLux field of view.

\textbf{I16555-0823 (GJ 644 C)}
While the star I16555-0823 (also known as VB~8) itself is single in the AstraLux sensitivity range, it is part of a higher-order multiple (at least triple, possibly quintuple) as a known 1500~AU companion to GJ~644, which was discussed in our \citet{janson2012} study.

\textbf{I18180+3846W (LHS 461)}
I18180+3846W has a 10\arcsec\ companion to the East registered in WDS. It is visible in the AstraLux images, but outside of the completeness range. There are no other candidates observed in the field.

\textbf{I18427+1354 (GJ 4071)}
There is a point source at 3.7\arcsec\ separation from I18427+1354. The system has only been observed in one epoch, but based on the colors of the candidate ($\Delta z^{\prime} = 5.8 \pm 0.1$~mag and $\Delta i^{\prime} = 5.7 \pm 0.1$~mag), it counts as a probable background contaminant in our analysis.

\textbf{I19312+3607 (G 125-15)}
While I19312+3607 has no candidates visible in the AstraLux images, it is actually a triple system. I19312+3607 itself is noted as a very close ($<$0.01~AU) double-lined spectroscopic binary in \citet{shkolnik2010}. Additionally, there is a wide common proper motion companion at 46\arcsec\ \citep{caballero2010}.

\textbf{I20298+0941 (HU Del)}
The I20298+0941 system is a well known astrometric binary \citep[e.g.][]{benedict2000}, but to our knowledge, the AstraLux images represent the first instance in which the binary has been spatially resolved.

\textbf{I20433+5520 (GJ 802 A)}
I20433+5520 is a well-studied close triple system \citep[e.g.][]{ireland2008}. It consists of a spectroscopic pair with a period of only 19~h, and a brown dwarf at a separation of $\sim$90~mas. Both are too close to be resolved with AstraLux, and there are no other candidates in the field of view.

\textbf{I21000+4004E (GJ 815 A)}
In addition to the component that we resolve with AstraLux, which was previously known and has been studied over a long timescale \citep[e.g.][]{lippincott1975}, the primary component of the resolved pair is a 3.3 day spectroscopic binary \citep{pourbaix2000},

\textbf{I21109+4657S (G 212-27)}
Both of the point sources in the AstraLux field of view are consistent with static background objects that do not share a common proper motion with the primay star, hence they are contaminants rather than physical companions. Their colors ($\Delta z^{\prime} = 4.0 \pm 0.3$~mag and $\Delta i^{\prime} = 3.3 \pm 0.5$~mag for the closer point source and $\Delta z^{\prime} = 4.8 \pm 0.3$~mag and $\Delta i^{\prime} = 4.3 \pm 0.1$~mag for the farther one) verify this conclusion.

\textbf{I21160+2951E (GJ 4185 A)}
Although I21160+2951E is single within the AstraLux field of view, it has a wide companion at 26\arcsec\ separation. It also appears that I21160+2951E is itself a close binary pair; this is implied in \citet{shkolnik2012}, but an as of yet unpublished paper is referred to for the specific properties of this pair. Since a general comment is made that separations down to 40~mas are being probed, it is presumably the case that the third component of the system is simply too close in to be resolved with AstraLux.

\textbf{I21376+0137 (J21374019+0137137)}
This newly discovered binary candidate is a probable member of the $\beta$~Pic moving group according to the \citet{schlieder2012b} study. Its relatively small projected separation of $\sim$4.5~AU implies that its orbit could be dynamically constrained in a reasonable timeframe, which makes it a potential benchmark binary in the future.

\textbf{I23318+1956E (EQ~Peg~B)}
This wide binary used to have a separation that would have kept it inside of the completeness range of AstraLux (e.g., 3.5\arcsec\ separation in 1941 according to WDS), but at the AstraLux epoch it is just outside of this range, at 5.4\arcsec. It therefore counts as being outside of this range.

{\scriptsize
\begin{longtable}{lllllllllllll}
\caption{General properties of all targets observed in the survey.}\\
\hline\hline
Lepine ID & $\alpha$  & $\delta$ & PM$_{\alpha}$ & PM$_{\delta}$ & $\pi$ & $e_{\pi}$ & Ref$^{\rm a}$ & SpT$^{\rm b}$ & Ref$^{\rm c}$ & $J$ & Mult$^{\rm d}$ & ND$^{\rm e}$ \\
 & (hh mm ss) & (dd mm ss) & (mas) & (mas) & (mas) & (mas) & & & & (mag) & & \\
\hline
\endfirsthead
\caption{continued.}\\
\hline\hline
Lepine ID & $\alpha$  & $\delta$ & PM$_{\alpha}$ & PM$_{\delta}$ & $\pi$ & $e_{\pi}$ & Ref$^{\rm a}$ & SpT$^{\rm b}$ & Ref$^{\rm c}$ & $J$ & Mult$^{\rm d}$ & ND$^{\rm e}$ \\
 & (hh mm ss) & (dd mm ss) & (mas) & (mas) & (mas) & (mas) & & & & (mag) & & \\
\hline
\endhead
\hline
\endfoot
\hline
\multicolumn{13}{l}{$^a$References for parallax. PHOT means photometric parallax from \citet{lepine2011} and G91 from \citet{gliese1991}, all other are trigonometric parallaxes.}\\
\multicolumn{13}{l}{C05: \citet{costa2005}, D14: \citet{dittman2014}, G08: \citet{gatewood2008}, G09: \citet{gatewood2009}, }\\
\multicolumn{13}{l}{H93: \citet{harrington1993}, H06: \citet{henry2006}, K10: \citet{khrutskaya2010}, L09: \citet{lepine2009}, M92: \citet{monet1992}, }\\
\multicolumn{13}{l}{R10: \citet{riedel2010}, Sh12: \citet{shkolnik2012}, vA95: \citet{vanaltena1995}, vL07: \citet{vanleeuwen2007}.}\\
\multicolumn{13}{l}{$^b$Spectral type; lower case notation implies a color-based SpT estimation which may be biased, see text.}\\
\multicolumn{13}{l}{$^c$References for spectral type. A00: \citet{alcala2000}, A09: \citet{agueros2009}, BS08: \citet{bender2008}, D98: \citet{delfosse1998}, }\\
\multicolumn{13}{l}{F09: \citet{faherty2009}, J14: \citet{jao2014}, J09: \citet{jenkins2009}, L09: \citet{lepine2009}, M13: \citet{mann2013}, }\\
\multicolumn{13}{l}{M03: \citet{mohanty2003}, M98: \citet{motch1998}, P91: \citet{prosser1991}, R95: \citet{reid1995}, R03: \citet{reid2003}, }\\
\multicolumn{13}{l}{R04: \citet{reid2004}, R07: \citet{reid2007}, S05: \citet{scholz2005}, Sch12: \citet{schlieder2012a}, Sh09: \citet{shkolnik2009}.}\\
\multicolumn{13}{l}{U: Unknown, spectral type listed in SIMBAD but it was not possible to locate the source.}\\
\multicolumn{13}{l}{$^d$Multiplicity flag. S: Single, as far as is known. M: Multiple within the AstraLux sensitivity range. C: A known companion exists in the}\\
\multicolumn{13}{l}{literature but is too close in for AstraLux. W: A known companion exists outside of the AstraLux completeness range. O: An object is}\\
\multicolumn{13}{l}{observed in the AstraLux image, but is outside of the completeness range. BG: One or more suspected or confirmed background objects are}\\
\multicolumn{13}{l}{observed in the images. See individual notes for detailed comments.}\\
\multicolumn{13}{l}{$^e$Flag for whether the companion is a new discovery (Y) ir not (N).}\\
\hline
\endlastfoot
I00066-0705	&	00 06 39.249	&	-07 05 35.33	&	-104	&	89	&	69.9	&	21.0	&	PHOT	&	M3.5	&	R07	&	9.83	&	M	&	N	\\
I00077+6022	&	00 07 42.620	&	+60 22 54.34	&	340	&	-27	&	68.6	&	2.0	&	D14	&	M3.8	&	Sh09	&	8.91	&	M	&	Y	\\
I00088+2050	&	00 08 53.922	&	+20 50 25.45	&	-65	&	-247	&	67.5	&	2.7	&	D14	&	M4.5	&	R95	&	8.87	&	M	&	N	\\
I00115+5908	&	00 11 31.808	&	+59 08 39.87	&	-918	&	-1164	&	108.3	&	1.4	&	L09	&	M6.5	&	L09	&	9.94	&	S	&	---	\\
I00132+6919N	&	00 13 15.850	&	+69 19 37.62	&	717	&	-292	&	49.9	&	6.0	&	vL07	&	M3.0	&	R95	&	8.56	&	M	&	N	\\
I00162+1951E	&	00 16 16.142	&	+19 51 50.61	&	704	&	-740	&	66.1	&	1.6	&	vA95	&	M4.0	&	R95	&	8.89	&	S	&	---	\\
I00169+0507	&	00 16 56.298	&	+05 07 26.54	&	-107	&	-629	&	56.9	&	3.7	&	vA95	&	M4.5	&	R95	&	9.40	&	S	&	---	\\
I00235+7711	&	00 23 31.836	&	+77 11 26.73	&	-848	&	26	&	52.0	&	2.0	&	vL07	&	M4.0	&	R95	&	9.93	&	O	&	---	\\
\end{longtable}
}

{\scriptsize
\begin{longtable}{lllllllll}
\caption{Astrometric properties of the binaries and background stars in the survey.}\\
\hline\hline
Lepine ID & Other ID & Pair & $\rho$ & $\theta$ & Epoch & Ref$^{\rm a}$ & CPM$^{\rm b}$ & OM$^{\rm c}$ \\
 & & & (\arcsec) & (deg) & (yr) &  &  & \\
\hline
\endfirsthead
\caption{continued.}\\
\hline\hline
Lepine ID & Other ID & Pair & $\rho$ & $\theta$ & Epoch & Ref$^{\rm a}$ & CPM$^{\rm b}$ & OM$^{\rm c}$ \\
 & & & (\arcsec) & (deg) & (yr) &  &  & \\
\hline
\endhead
\hline
\endfoot
\hline
\multicolumn{9}{l}{$^a$Source of the astrometry at the given epoch. Only one archival epoch listed per target.}\\
\multicolumn{9}{l}{TP: This paper.}\\
\multicolumn{9}{l}{WDS: Washington Double Star Catalog, \citet{mason2001}.}\\
\multicolumn{9}{l}{M01: \citet{mccarthy2001}.}\\
\multicolumn{9}{l}{B04: \citet{beuzit2004}; uniform errors assumed.}\\
\multicolumn{9}{l}{M06: \citet{montagnier2006}.}\\
\multicolumn{9}{l}{P06: \citet{pravdo2006}.}\\
\multicolumn{9}{l}{D07: \citet{daemgen2007}.}\\
\multicolumn{9}{l}{L08: \citet{law2008}.}\\
\multicolumn{9}{l}{D10: \citet{dupuy2010} orbital analysis; original data point from \citet{montagnier2006}.}\\
\multicolumn{9}{l}{H12: \citet{horch2012}.}\\
\multicolumn{9}{l}{J12: \citet{janson2012}.}\\
\multicolumn{9}{l}{J13: \citet{jodar2013}.}\\
\multicolumn{9}{l}{$^b$Flag for common proper motion (evaluated between the first and last listed epochs), yes (Y), no (N) or inferred (I).}\\
\multicolumn{9}{l}{$^c$Flag for orbital motion (between the first and last listed epochs), yes (Y), no (N), unclear (U), or background (BG).}\\
\multicolumn{9}{l}{$^d$Quoted angle assumes a phase shift, see individual note for the target.}\\
\hline
\endlastfoot
I00066-0705	&	---	&	AB	&	0.230$\pm$0.006	&	6.5$\pm$0.5	&	2008.63	&	J12	&	Y	&	Y	\\
\ I00066-0705	&	---	&	AB	&	0.322$\pm$0.004	&	1.9$\pm$0.3	&	2012.65	&	TP	&		&		\\
\ I00066-0705	&	---	&	AB	&	0.337$\pm$0.003	&	1.4$\pm$0.3	&	2012.90	&	TP	&		&		\\
I00077+6022	&	G 217-32	&	AB	&	0.612$\pm$0.006	&	82.5$\pm$0.3	&	2011.85	&	TP	&	Y	&	Y	\\
\ I00077+6022	&	G 217-32	&	AB	&	0.661$\pm$0.007	&	86.9$\pm$0.3	&	2012.65	&	TP	&		&		\\
\ I00077+6022	&	G 217-32	&	AB	&	0.674$\pm$0.007	&	87.9$\pm$0.3	&	2012.89	&	TP	&		&		\\
I00088+2050	&	GJ 3010	&	AB	&	0.111$\pm$0.005	&	169.9$\pm$0.5	&	2001.60	&	B04	&	Y	&	Y	\\
\ I00088+2050	&	GJ 3010	&	AB	&	0.133$\pm$0.005	&	271.9$\pm$1.7	&	2012.02	&	TP	&		&		\\
\end{longtable}
}

{\scriptsize
\begin{longtable}{lllllllllllll}
\caption{Photometric and physcial properties of the binaries in the survey.}\\
\hline\hline
Lepine ID & Other ID & Pair & $\Delta z^{\prime}$  & $\Delta i^{\prime}$ & $\tau_{\rm low}$ & $\tau_{\rm high}$ & Ref$^{\rm a}$ & $m_{\rm A}$ & $m_{\rm B}$ & $q$ & $a_{\rm est}$ & SC$^{\rm b}$ \\
 & & & (mag) & (mag) & (Myr) & (Myr) &  & ($M_{\rm Sun}$) & ($M_{\rm Sun}$) & & (AU) & \\
\hline
\endfirsthead
\caption{continued.}\\
\hline\hline
Lepine ID & Other ID & Pair & $\Delta z^{\prime}$  & $\Delta i^{\prime}$ & $\tau_{\rm low}$ & $\tau_{\rm high}$ & Ref$^{\rm a}$ & $m_{\rm A}$ & $m_{\rm B}$ & $q$ & $a_{\rm est}$ & SC$^{\rm b}$ \\
 & & & (mag) & (mag) & (Myr) & (Myr) &  & ($M_{\rm Sun}$) & ($M_{\rm Sun}$) & & (AU) & \\
\hline
\endhead
\hline
\endfoot
\hline
\multicolumn{13}{l}{$^a$Basis for the age estimation.}\\
\multicolumn{13}{l}{Sc12: Member of AB~Dor of $\beta$~Pic moving group according to \citet{schlieder2012b}.}\\
\multicolumn{13}{l}{Sh12: Estimated age in \citet{shkolnik2012}.}\\
\multicolumn{13}{l}{M13: Member of a young moving group according to \citet{malo2013}.}\\
\multicolumn{13}{l}{VY: Assumed very young based on high X-ray flux \citep[as listed in][]{lepine2011}.}\\
\multicolumn{13}{l}{MY: Assumed moderately young based on moderate X-ray flux.}\\
\multicolumn{13}{l}{NY: Assumed not young based on lack of detectable X-ray flux.}\\
\multicolumn{13}{l}{$^b$Flag for whether targets is `statistically clean' (Y) or not (N), see Sect. \ref{s:multfrac}.}\\
\hline
\endlastfoot
I00066-0705	&	---	&	AB	&	1.15$\pm$0.05	&	1.43$\pm$0.04	&	1000	&	10000	&	NY	&	0.35$\pm$0.09	&	0.21$\pm$0.05	&	0.61$\pm$0.05	&	5.6	&	N	\\
I00077+6022	&	G 217-32	&	AB	&	0.72$\pm$0.12	&	0.86$\pm$0.16	&	35	&	300	&	Sh12	&	0.17$\pm$0.07	&	0.12$\pm$0.06	&	0.68$\pm$0.06	&	9.8	&	Y	\\
I00088+2050	&	GJ 3010	&	AB	&	1.20$\pm$0.07	&	1.59$\pm$0.10	&	30	&	300	&	VY	&	0.20$\pm$0.09	&	0.11$\pm$0.06	&	0.52$\pm$0.05	&	2.0	&	Y	\\
I00132+6919N	&	GJ 11 B	&	AB	&	0.81$\pm$0.01	&	0.69$\pm$0.02	&	1000	&	10000	&	NY	&	0.38$\pm$0.10	&	0.26$\pm$0.06	&	0.69$\pm$0.02	&	17.2	&	Y	\\
I00395+1454N	&	G 32-37 B	&	AB	&	1.12$\pm$0.10	&	---	&	300	&	1000	&	MY	&	0.33$\pm$0.03	&	0.19$\pm$0.01	&	0.58$\pm$0.01	&	4.3	&	N	\\
I00489+4435	&	GJ 3058	&	AB	&	0.28$\pm$0.01	&	0.35$\pm$0.02	&	50	&	150	&	Sc12	&	0.22$\pm$0.10	&	0.18$\pm$0.08	&	0.84$\pm$0.01	&	19.0	&	Y	\\
I01032+7113	&	LHS 1182	&	AB	&	1.47$\pm$0.06	&	1.62$\pm$0.07	&	1000	&	10000	&	NY	&	0.21$\pm$0.04	&	0.13$\pm$0.02	&	0.59$\pm$0.05	&	2.7	&	Y	\\
I01114+1526	&	GJ 3076	&	AB	&	1.71$\pm$0.86	&	1.46$\pm$0.15	&	10	&	20	&	M13	&	0.10$\pm$0.03	&	0.04$\pm$0.01	&	0.45$\pm$0.06	&	5.6	&	Y	\\
\end{longtable}
}

\clearpage

\end{document}